\documentclass[final,5p,times,review]{elsarticle}

\usepackage{graphicx}
\usepackage{dcolumn}
\usepackage{bm}
\usepackage{hyperref}

\newcommand{\cdummy}{\cdot}

\newcommand{\tmmathbf}[1]{\ensuremath{\boldsymbol{#1}}}

\def\etal{{\em{et al }}}
\begin{document}


\title{Investigating elastic constants across diverse strain-matrix sets}

\author[1,2]{Zhong-Li Liu}
\author[1]{Ya-Dong Wei}
\author[1]{Xiao-Dong Xu}
\author[1]{Wei-Qi Li}
\author[1]{Gang Lv}
\author[1]{Jian-Qun Yang}
\author[1]{Xing-Ji Li}
\ead{lxj0218@hit.edu.cn}
\author[3]{Chinedu E. Ekuma}
\ead{che218@lehigh.edu}

\address[1]{Technology Innovation Center of Materials and Devices at Extreme Environment, School of Materials Science and Engineering, Harbin Institute of Technology, Harbin, China}
\address[2]{College of Physics and Electric Information, Luoyang Normal University, Luoyang 471934, China}

\address[3]{Department of Physics, Lehigh University, Bethlehem, PA 18015 USA}

\date{\today}

\begin{abstract}
Elastic constants and mechanical properties play a pivotal role across multiple disciplines and engineering applications. We introduced the optimized high-efficient strain-matrix set (OHESS) that determines the second-order elastic constants of materials using the stress-strain method. Herein, we systematically investigate the computational efficiency of  OHESS across a broad range of crystal systems and compare it with other notable stress-strain approaches, such as the single-element strain-matrix sets and the universal linear-independent coupling strains. Notably, our data affirm the superior efficacy of OHESS among the strain-matrix sets under consideration. We believe OHESS will markedly improve computational efficiency in determining the elastic constants and mechanical properties, becoming an indispensable tool for material research, design, and high-throughput screening
\end{abstract}

\maketitle

\section{Introduction}
Elastic constants serve as fundamental metrics that shed light on the physical and chemical attributes of crystalline materials. These constants elucidate the mechanical behavior of materials when subjected to diverse external forces and stresses, offering insights into their inherent strength and potential hardness (See Ref~\cite{Tehrani2019} and
references therein). The correlation between the velocities of elastic waves and elastic constants is profound, paving the way for understanding systematic acoustic properties in various materials. Specifically, for mineral materials, these elastic constants provide critical information for understanding the characteristics of the seismic wave as it travels through the Earth's interior~\cite{Lay1995}. Furthermore, elastic constants hold the key to elucidating the thermodynamic properties of materials. They can guide us in analyzing aspects such as the phonon dispersion relation, thermal fluctuations, Debye temperature, the Gr{\"u}neisen parameter, and even the melting point~\cite{Ashcroft1976}. On the atomic scale, these constants act as mirrors reflecting the strength of chemical bonds in various orientations within materials~\cite{Tatsumi2003}. Consequently, the role of elastic constants extends across diverse fields, including but not limited to physics, condensed matter physics, materials science, geophysics, chemistry, and various engineering disciplines.

Historically, the determination of elastic constants predominantly rested on experimental methods. However, due to the inherent challenges of these methods, the elastic constants for some materials remain elusive in the experimental literature. Advancements in computational techniques such as density functional theory (DFT)~\cite{Hohenberg1964,Kohn1965} have enabled the determination of the elastic constants of diverse materials with remarkable accuracy. Computational approaches prove invaluable when venturing into high-pressure regimes, where traditional experiments cannot  directly measure elastic constants. In such scenarios, DFT-based calculations become paramount, playing a dual role: unearthing the nuances of material properties and aiding in the conceptualization of novel materials.

A key application of computed elastic constants is in the realm of new material design, specifically in the domain of crystal structure prediction {\cite{Liu2014,Wang2010,Wang2012,Glass2006,Pickard2011,Lonie2011}}. Here, they serve a dual purpose. Firstly, they are employed to assess the stability of proposed structures, often benchmarked against the Born elastic stability criteria {\cite{Wu2007,Mouhat2014,Nye1985,Born1954,Liu2017,Yu2010}}. Secondly, these constants find a coveted place in the vast reservoir of computational materials databases, notably the Materials Project (MP) database~\cite{Jong2015}. Such databases, underpinned by the principles of materials informatics, serve as foundational platforms for driving innovative design of new materials.

Elastic constants can be computed via two primary methods: the strain-energy method and the strain-stress method.~\cite{Wang2021,Golesorkhtabar2013,Perger2009}

The strain-stress method, in contrast to the strain-energy approach, hinges on the availability of highly accurate stress tensors and typically requires fewer strain sets.~\cite{Liu2022} One pivotal consideration when employing the strain-stress technique is the need for denser K-point meshes, ensuring the precision of the stress tensors. While this introduces a higher computational overhead, the approach is favored for its simplicity in implementation. Another compelling advantage of the strain-stress method is its directness when accounting for pressure-dependent effects. In particular, it allows for the straightforward calculation of a material's elastic constants without the need for intricate pressure corrections. This contrasts with the strain-energy method, where pressure corrections can be complex and demanding.

Expanding upon the earlier methodology, Yu \etal introduced the concept of universal linear-independent coupling strains (ULICS) within the stress-strain framework {\cite{Yu2010}}. In this approach, various stress components are intricately coupled to collectively determine the complete set of elastic constants. A notable implication of deploying multiple strain components in ULICS is the potential reduction in the symmetry of the strained crystal, which, in turn, can substantially increase the computational cost. Several other approaches to computationally obtain the elastic constant of materials have been implemented in a number of previous works.~\cite{Perger2009,Yu2010,Golesorkhtabar2013,Wang2021} However, accurate and efficient calculation of the elastic constant of materials is essential, especially in new materials design and high-throughput materials screening. There is a need to explore the most efficient computational approach to obtain the elastic constants of materials. In this work, we explore and present the efficiency of various strain matrix sets applicable across various crystal systems, encompassing both three-dimensional (3D) and two-dimensional (2D) structures. Our optimized sets aim to preserve the inherent symmetry of the crystal during deformation, which considerably improves the computational efficiency.

The structure of this paper is as follows: After the introduction, Section \ref{method} outlines the theory behind calculating elastic constants and the optimization of strain matrix sets. Section \ref{result} evaluates and compares the computational accuracy and efficiency of various strain matrix sets. Our discussions and conclusions are summarized in Sec. \ref{conclusion}.

\section{Methodology}\label{method}

\subsection{ Elasticity theory}
The study of a material's elastic properties provides insight into its response to externally applied loads. Elastic properties dictate how a material deforms under these loads, which can be essential for various applications, from engineering to materials design. According to Hooke's law, within the linear elastic regime, the stresses denoted as \( \sigma_i \), in a crystal are directly proportional to the applied strains, \( \varepsilon_j \),
\begin{equation}
\sigma_i = \sum_{j = 1}^6 C_{i j} \varepsilon_j, \label{hooke}
\end{equation}
where \( C_{i j} \) is the elastic stiffness constants or the coefficients that govern the proportionality. These constants, known as the components of the elastic tensor, play a pivotal role in characterizing the elastic behavior of the material. Equation (\ref{hooke}) employs the Voigt notation {\cite{Voigt1928}}, where indices 1, 2, ..., and 6 correspond to the stress or strain components \( xx, yy, zz, yz, zx, \) and \( xy \), respectively. This compact notation simplifies tensor representations, particularly for isotropic and cubic materials. In determining these elastic constants computationally, one applies specific strain sets to a crystal and then calculates the resulting stress tensors. These calculations can be performed using empirical atomic potential methods, which offer rapid approximations, or more precise first-principles methods based on quantum mechanics, which provide a more accurate picture at the expense of higher computational cost.

For 3D materials, the strain matrix, in Voigt notation, can be represented as:
\begin{equation}
\tmmathbf{\varepsilon}= \left[ \begin{array}{l}
\varepsilon_1 \quad \varepsilon_6/2 \quad \varepsilon_5/2\\
\varepsilon_6/2 \quad \varepsilon_2 \quad \varepsilon_4/2\\
\varepsilon_5/2 \quad \varepsilon_4/2 \quad \varepsilon_3
\end{array} \right] .
\end{equation}
In this representation, the off-diagonal terms account for the shear components of the strain, and they are halved due to the symmetrical nature of the stress and strain tensors. For 2D layered materials, one typically assumes the crystal plane lies in the \(x y\) plane. Given this, the strain matrix for 2D materials can be significantly simplified compared to the 3D scenario and can be represented as:
\begin{equation}
\tmmathbf{\varepsilon}= \left[ \begin{array}{l}
\varepsilon_1 \quad \varepsilon_6/2 \quad 0\\
\varepsilon_6/2 \quad \varepsilon_2 \quad 0\\
0 \qquad 0 \quad 0
\end{array} \right] .
\end{equation}
To account for the deformation induced in the crystal, we define the deformation matrix, \( \tmmathbf{D} \)
\begin{equation}
\tmmathbf{D}=\tmmathbf{I}+\tmmathbf{\varepsilon},
\end{equation}
where \( \tmmathbf{I} \) represents a 3$\times$3 identity matrix. The deformed crystal lattice vectors can then be expressed as:
\begin{equation}
\tmmathbf{A}' =\tmmathbf{A} \cdummy \tmmathbf{D},
\end{equation}
where \( \tmmathbf{A} \) denotes the original, undeformed lattice vectors.

Crucially, various crystal systems exhibit different numbers of independent second-order elastic constants (SOECs), which arise from their inherent lattice symmetries. The full elastic constant set consists of 21 for 3D materials and 6 for 2D materials. However, symmetries in the lattice can reduce these numbers, making the calculations more tractable. The specific numbers of SOECs for different lattice symmetries are summarized in Table \ref{num}. With a thorough understanding of these SOECs and their symmetries, one can then tailor the strain matrix sets accordingly. By doing so, the computational efficiency of deriving the elastic constants can be enhanced, paving the way for quicker insights into the mechanical properties of materials.

\begin{table*}[htp!]
	\caption{Summarize of the number of second-order elastic constants (SOECs) for various lattice symmetries in both 3D and 2D structures. 'SGN' denotes the space group number associated with specific crystal structures.\label{num}}
	\centering{
		\begin{tabular}{clccl}
		\hline
			Dimensional & Crystal system &  SGN &  Number of matrices  & Prototype\\
			\hline
			3D & Cubic & 195-230 & 3 & C, Al, CsCl\\
			& Hexagonal & 168-194 & 5 & Os, Ti, TiB$_2$\\
			& Rhombohedral I & 149-167 & 6 & Al$_2$O$_3$\\
			& Rhombohedral II & 143-148 & 7 & CaMg(CO$_3$)$_2$\\
			& Tetragonal I & 89-142 & 6 & MgF$_2$\\
			& Tetragonal II & 75-88 & 7 & CaMoO$_4$\\
			& Orthorhombic & 16-74 & 9 & TiSi$_2$\\
			& Monoclinic & 3-15 & 13 & ZrO$_2$\\
			& Triclinic & 1-2 & 21 & ReS$_2$\\
			2D & Hexagonal & - & 2 & Graphene, MoS$_2$\\
			& Square & - & 3 & FeSe\\
			& Rectangular & - & 4 & Phosphorene, AuSe\\
			& Oblique & - & 6 & -\\
		\hline
		\end{tabular}}

\end{table*}

\subsection{Optimization of strain matrix sets}
Yu \etal introduced the ULICS, a novel approach designed to couple all stress components, thereby facilitating the simultaneous extraction of the complete set of elastic constants~\cite{Yu2010}. For a deeper dive into the intricacies of the ULICS, we recommend consulting Ref. \cite{Yu2010}. However, our extensive testing revealed that while ULICS offers several advantages, it also has certain drawbacks. Notably, the deformations it induces on crystal lattice vectors can inadvertently lower their symmetries. This phenomenon presents challenges, especially when the density functional theory (DFT) method is employed, as it leads to an increase in computational cost and inevitably a potential wrong prediction of the elastic constants. More details on these challenges and our findings can be found in Section \ref{rd}.

To address the shortcomings of ULICS and reduce computational overhead, we developed the Optimal High Efficiency Strain-Matrix Sets (OHESS). This innovative approach primarily focuses on preserving the inherent symmetries of crystals. It achieves this by optimizing the strain matrix sets, ensuring the computational process remains efficient without compromising on accuracy. Our tests, detailed in Section \ref{rd}, highlight that OHESS not only meets, but surpasses the calculation efficiency of ULICS. For practical insights, we have cataloged the OHESS for a range of lattice systems, including both 3D and 2D structures, in Table \ref{ohess}. This table illustrates how specific strain sets can be used to deduce certain elastic constants.

\begin{table*}[htp!]
		\caption{The complete elastic constant parameters within the OHESS approach for different lattice systems of 3D and 2D crystals.\label{ohess}}
\centering{
	\begin{tabular}{cccccccccl}
		\hline
		Dimensional & Lattice system & Number of matrices & $\epsilon_1$ &
		$\epsilon_2$ & $\epsilon_3$ & $\epsilon_4$ & $\epsilon_5$ & $\epsilon_6$ &
		Derived $C_{ij}$\\
		\hline
		3D & Cubic & 1 & $\delta$ & 0 & 0 & $\delta$ & 0 & 0 & $C_{11}$, $C_{12}$,
		$C_{44}$\\
		& Hexagonal & 2 & $\delta$ & 0 & 0 & 0 & 0 & 0 & $C_{11}$, $C_{12}$,
		$C_{13}$\\
		&  &  & 0 & 0 & $\delta$ & $\delta$ & 0 & 0 & $C_{33}$, $C_{44}$\\
		& Rhombohedral I & 2 & $\delta$ & 0 & 0 & 0 & 0 & 0 & $C_{11}$, $C_{12}$,
		$C_{13}$, $C_{14}$\\
		&  &  & 0 & 0 & $\delta$ & $\delta$ & 0 & 0 & $C_{33}$, $C_{44}$\\
		& Rhombohedral II & 2 & $\delta$ & 0 & 0 & 0 & 0 & 0 & $C_{11}$,
		$C_{12}$, $C_{13}$, $C_{14}$, $C_{15}$\\
		&  &  & 0 & 0 & $\delta$ & $\delta$ & 0 & 0 & $C_{33}$, $C_{44}$\\
		& Tetragonal I & 2 & $\delta$ & 0 & 0 & 0 & 0 & 0 & $C_{11}$, $C_{12}$,
		$C_{13}$\\
		&  &  & 0 & 0 & $\delta$ & $\delta$ & 0 & 0 & $C_{33}$, $C_{44}$\\
		& Tetragonal II & 2 & $\delta$ & 0 & 0 & 0 & 0 & 0 & $C_{11}$, $C_{12}$,
		$C_{13}$, $C_{16}$\\
		&  &  & 0 & 0 & $\delta$ & $\delta$ & 0 & 0 & $C_{33}$, $C_{44}$\\
		& Orthorhombic & 3 & $\delta$ & 0 & 0 & 0 & 0 & 0 & $C_{11}$, $C_{12}$,
		$C_{13}$\\
		&  &  & 0 & $\delta$ & 0 & 0 & 0 & 0 & $C_{22}$, $C_{23}$\\
		&  &  & 0 & 0 & $\delta$ & $\delta$ & $\delta$ & $\delta$ & $C_{33}$,
		$C_{44}$, $C_{55}$, $C_{66}$\\
		& Monoclinic & 4 & $\delta$ & 0 & 0 & 0 & 0 & 0 & $C_{11}$, $C_{12}$,
		$C_{13}$, $C_{16}$\\
		&  &  & 0 & $\delta$ & 0 & 0 & 0 & 0 & $C_{22}$, $C_{23}$, $C_{26}$\\
		&  &  & 0 & 0 & $\delta$ & $\delta$ & 0 & 0 & $C_{33}$, $C_{36}$,
		$C_{44}$, $C_{45}$\\
		&  &  & 0 & 0 & 0 & 0 & $\delta$ & $\delta$ & $C_{55}$, $C_{66}$\\
		& Triclinic & 6 & $\delta$ & 0 & 0 & 0 & 0 & 0 & $C_{11}$, $C_{12}$,
		$C_{13}$, $C_{14}$, $C_{15}$, $C_{16}$\\
		&  &  & 0 & $\delta$ & 0 & 0 & 0 & 0 & $C_{22}$, $C_{23}$, $C_{24}$,
		$C_{25}$, $C_{26}$\\
		&  &  & 0 & 0 & $\delta$ & 0 & 0 & 0 & $C_{33}$, $C_{34}$, $C_{35}$,
		$C_{36}$\\
		&  &  & 0 & 0 & 0 & $\delta$ & 0 & 0 & $C_{44}$, $C_{45}$, $C_{46}$\\
		&  &  & 0 & 0 & 0 & 0 & $\delta$ & 0 & $C_{55}$, $C_{56}$\\
		&  &  & 0 & 0 & 0 & 0 & 0 & $\delta$ & $C_{66}$\\
		2D & Hexagonal & 1 & $\delta$ & 0 & 0 & 0 & 0 & $\delta$ & $C_{11}$,
		$C_{12}$\\
		& Square & 1 & $\delta$ & 0 & 0 & 0 & 0 & $\delta$ & $C_{11}$, $C_{12}$,
		$C_{66}$\\
		& Rectangular & 2 & $\delta$ & 0 & 0 & 0 & 0 & $\delta$ & $C_{11}$,
		$C_{12}$, $C_{66}$\\
		&  &  & 0 & $\delta$ & 0 & 0 & 0 & 0 & $C_{22}$\\
		& Oblique & 3 & $\delta$ & 0 & 0 & 0 & 0 & 0 & $C_{11}$, $C_{12}$,
		$C_{16}$\\
		&  &  & 0 & $\delta$ & 0 & 0 & 0 & 0 & $C_{22}$, $C_{26}$\\
		&  &  & 0 & 0 & 0 & 0 & 0 & $\delta$ & $C_{33}$\\
		\hline
	\end{tabular}}

\end{table*}

We also evaluated the All Single-Element Strain-Matrix Sets (ASESS) as a benchmark. Distinct from ULICS, ASESS opts to decouple the stress components during calculations, offering a different computational approach. A comprehensive outline of the ASESS and the elastic constants derived from it is presented in Table \ref{asess}. Further, we conducted a comparative analysis detailing the number of strain matrix sets utilized by ASESS, OHESS, and ULICS. This is tabulated in Table~\ref{comp}. An intriguing observation from this comparison is that OHESS and ULICS have almost identical counts for most structures. However, a notable exception arises with the monoclinic structure, where OHESS demonstrates efficiency by using fewer strain matrix sets than ULICS.

\begin{table*}[htp!]
	\caption{The complete elastic constant parameters within the ASESS approach for different lattice systems of 3D and 2D crystals. \label{asess}}
\centering{
	\begin{tabular}{ccccccccl}
		\hline
		System & Number of matrices & $\epsilon_1$ & $\epsilon_2$ &
		$\epsilon_3$ & $\epsilon_4$ & $\epsilon_5$ & $\epsilon_6$ & Derived
		$C_{ij}$\\
		\hline
		Any 3D & 6 & $\delta$ & 0 & 0 & 0 & 0 & 0 & $C_{11}$, $C_{12}$, $C_{13}$,
		$C_{14}$, $C_{15}$, $C_{16}$\\
		&  & 0 & $\delta$ & 0 & 0 & 0 & 0 & $C_{22}$, $C_{23}$, $C_{24}$,
		$C_{25}$, $C_{26}$\\
		&  & 0 & 0 & $\delta$ & 0 & 0 & 0 & $C_{33}$, $C_{34}$, $C_{35}$,
		$C_{36}$\\
		&  & 0 & 0 & 0 & $\delta$ & 0 & 0 & $C_{44}$, $C_{45}$, $C_{46}$\\
		&  & 0 & 0 & 0 & 0 & $\delta$ & 0 & $C_{55}$, $C_{56}$\\
		&  & 0 & 0 & 0 & 0 & 0 & $\delta$ & $C_{66}$\\
		Any 2D & 3 & $\delta$ & 0 & 0 & 0 & 0 & 0 & $C_{11}$, $C_{12}$, $C_{16}$\\
		&  & 0 & $\delta$ & 0 & 0 & 0 & 0 & $C_{22}$, $C_{26}$\\
		&  & 0 & 0 & 0 & 0 & 0 & $\delta$ & $C_{66}$\\
		\hline
	\end{tabular}}

\end{table*}

\begin{table}[htp!]
	\caption{Comparison of the numbers of strain matrix sets used for ASESS, OHESS, and ULICS calculations.\label{comp}}

	\begin{tabular}{ccccc}
		\hline
		Dimensional & Lattice system & ASESS & OHESS & ULICS\\
		\hline
		3D & Cubic & 6 & 1 & 1\\
		& Hexagonal & 6 & 2 & 2\\
		& Rhombohedral & 6 & 2 & 2\\
		& Tetragonal & 6 & 2 & 2\\
		& Orthorhombic & 6 & 3 & 3\\
		& Monoclinic & 6 & 4 & 5\\
		& Triclinic & 6 & 6 & 6\\
		2D & Hexagonal & 3 & 1 & 1\\
		& Square & 3 & 1 & 1\\
		& Rectangular & 3 & 2 & 2\\
		& Oblique & 3 & 3 & 3\\
		\hline
	\end{tabular}

\end{table}

\subsection{The calculation details of stress tensors}
To ensure the accurate determination of elastic constants, we first undertook a meticulous procedure to optimize each crystal structure at ambient pressure. Following this optimization, the atomic positions were subjected to specific deformations as per the OHESS, ASESS, or ULICS methods. Under these deformations, the stress components were determined with high precision, with the forces exerted on each atom and the energy converging to a threshold of 0.001 eV/{\r A} and $10^{-6}$ eV, respectively. All structural optimizations and subsequent stress computations in this study were executed employing the projector augmented wave (PAW) method~\cite{Blochl1994} as implemented in the VASP electronic structure code~\cite{Kresse1999,Kresse1996}. We used the Perdew, Becke, and Ernzerhof (PBE)~\cite{Perdew1997} generalized gradient approximation (GGA) for the exchange-correlation functional across all calculations. 

In the context of strain matrix sets, we maintained consistent parameters across the OHESS, ULICS, and ASESS for elastic constant calculations. All the implementation has been incorporated in the ElasTool package \cite{Liu2022}. For the precise derivation of elastic constants, specific values of $\delta$, namely $-0.06$, $-0.03$, $0.0$, $0.03$, and $0.06$, were adopted within the strain matrix sets. These values were used for the first-order polynomial fitting, in line with Eq.\ref{hooke}. Notably, the same values were also integrated within the ULICS, marking them as the largest strain values, thereby ensuring a level comparison platform. An interesting observation to note, from Yu \etal~\cite{Yu2010}, is that the relatively small $\delta$ values in the order of $10^{-3}$ might not always be suitable due to the potential for numerical noise in certain scenarios.

\section{Results and discussions\label{rd}}\label{result}
Employing the three distinct strain sets: OHESS, ASESS, and ULICS, we meticulously optimized a collection of prototype materials, as enumerated in Table \ref{num}. These prototype materials span a variety of crystal systems, inclusive of both 3D and 2D structures. Our findings, which will be detailed in the subsequent subsection \ref{IIIA}, reveal that the OHESS, ASESS, and ULICS all demonstrate reasonable accuracy in deducing the elastic constants of the mentioned prototype materials. Notably, while each strain set displayed its unique strengths, the OHESS consistently emerged as the most efficient, exceeding both the ASESS and ULICS in terms of computational performance. This comparative assessment of efficiency is elaborated further in subsection \ref{IIIB}. In the following two Subsections, our focus will be on providing an in-depth comparative analysis of the accuracy and efficiency of the three strain sets in both 3D and 2D crystal systems.

\subsection{Comparison of elastic constants from different strain
	sets}\label{IIIA}

We undertook a comprehensive analysis of the elastic constants for a diverse array of prototype materials, as delineated in Table~\ref{num}. For our calculations, we employed three distinct strain sets: the OHESS, the ULICS, and the ASESS. The primary objective was to discern their respective accuracy across varying crystal systems. The 3D prototype materials listed in Table~\ref{num} encompass a wide spectrum of crystal systems, including the cubic, hexagonal, rhombohedral, tetragonal, orthorhombic, monoclinic, and triclinic systems. Moreover, these prototype materials are representative of a broad classification of crystal types—spanning ionic and covalent crystals, as well as metals. To ensure a holistic understanding, our evaluation was not confined merely to 3D materials; it was further extended to encompass 2D materials.

\subsubsection{3D materials}
We embarked on an assessment of the computational accuracy of the three strain sets - OHESS, ULICS, and ASESS - in determining the elastic constants for 3D crystals. Our examination spanned systems of varied symmetries, starting from the high-symmetry cubic system, and extending to the more complex low-symmetry monoclinic system. For the cubic system, we selected representative materials: diamond, Al, and CsCl. Their calculated elastic constants are presented in Table~\ref{tab-cubic}. These values are juxtaposed with data from the Materials Project~\cite{Jong2015} and relevant experimental findings. It is evident that the computed elastic constants, irrespective of the strain set employed, exhibited remarkable agreement with both the MP data \cite{Jong2015} and experimental measurements \cite{McSkimin1972,Jong2015,Slagle1967}. Notably, the precision of the OHESS results mirrors that of the ULICS and ASESS, highlighting its reliability and efficacy.

\begin{table}[htp!]
	\caption{The elastic constants calculated by the OHESS, ULICS, and ASESS approaches for the cubic prototype systems, in comparison with data from Materials Project and experiment.\label{tab-cubic}}

	\begin{tabular}{lllll}
		\hline
		System & Method & $C_{11}$ & $C_{12}$ & $C_{44}$\\
		\hline
		Diamond & OHESS & 1055.0 & 136.6 & 567.8\\
		& ULICS & 1063.4 & 145.0 & 582.1\\
		& ASESS & 1054.2 & 131.3 & 566.3\\
		& MP{\cite{Jong2015}} & 1054 & 126 & 562\\
		& Exp.{\cite{McSkimin1972}} & 1077.0 & 124.6 & 577.0\\
		Al & OHESS & 115.3 & 62.1 & 36.6\\
		& ULICS & 104.6 & 70.0 & 34.2\\
		& ASESS & 114.1 & 62.1 & 31.6\\
		& MP{\cite{Jong2015}} & 104 & 73 & 32\\
		& Exp.{\cite{Hazen1985}} & 108.0 & 62.0 & 28.3\\
		CsCl & OHESS & 33.4 & 5.9 & 5.5\\
		& ULICS & 33.4 & 6.7 & 6.4\\
		& ASESS & 33.0 & 5.5 & 5.0\\
		& MP{\cite{Jong2015}} & 34 & 6 & 5\\
		& Exp.{\cite{Slagle1967}} & 36.4 & 8.8 & 8.0\\
		\hline
	\end{tabular}

\end{table}

\begin{table}[htp!]
	\caption{The elastic constants calculated by the OHESS, ULICS, and ASESS approaches for the hexagonal prototype systems, in comparison with data from Materials Project and experiment.\label{hcp}}

		\begin{tabular}{lllllll}
			\hline
			System & Method & $C_{11}$ & $C_{12}$ & $C_{13}$ & $C_{33}$ & $C_{44}$\\
			\hline
			Os & OHESS & 768.8 & 238.9 & 231.4 & 858.4 & 265.4\\
			& ULICS & 747.9 & 243.3 & 238.2 & 855.4 & 258.8\\
			& ASESS & 768.8 &  241.3 & 225.2 & 850.1 & 256.7\\
			& MP{\cite{Jong2015}} & 730 & 226 & 220 & 824 & 252\\
			& Exp.{\cite{Pantea2009}} & 763.3 & 227.9 & 218.0 & 843.2 & 269.3\\
			Ti & OHESS & 172.8 & 92.3 & 84.4 & 189.4 & 38.2\\
			& ULICS & 173.2 & 91.5 & 84.9 & 185.0 & 38.5\\
			& ASESS & 172.8 & 91.9 & 84.5 & 191.8 & 39.4\\
			& MP{\cite{Jong2015}} & 177 & 83 & 76 & 191 & 42\\
			& Exp.{\cite{Smithells1983}} & 160 & 90 & 66 & 181 & 46\\
			TiB$_2$ & OHESS & 660.8 & 77.7 & 120.1 & 473.6 & 267.3\\
			& ULICS & 658.4 & 80.5 & 125.1 & 474.9 & 264.5\\
			& ASESS & 660.8 & 77.1 & 116.9 & 470.1 & 261.4\\
			& MP{\cite{Jong2015}} & 642 & 75 & 106 & 443 & 258\\
			& Exp.{\cite{Spoor1997}} & 660 & 48 & 93 & 432 & 260\\
			\hline
		\end{tabular}
\end{table}

\begin{figure}[htp!]
	\resizebox{8.5cm}{!}{\includegraphics{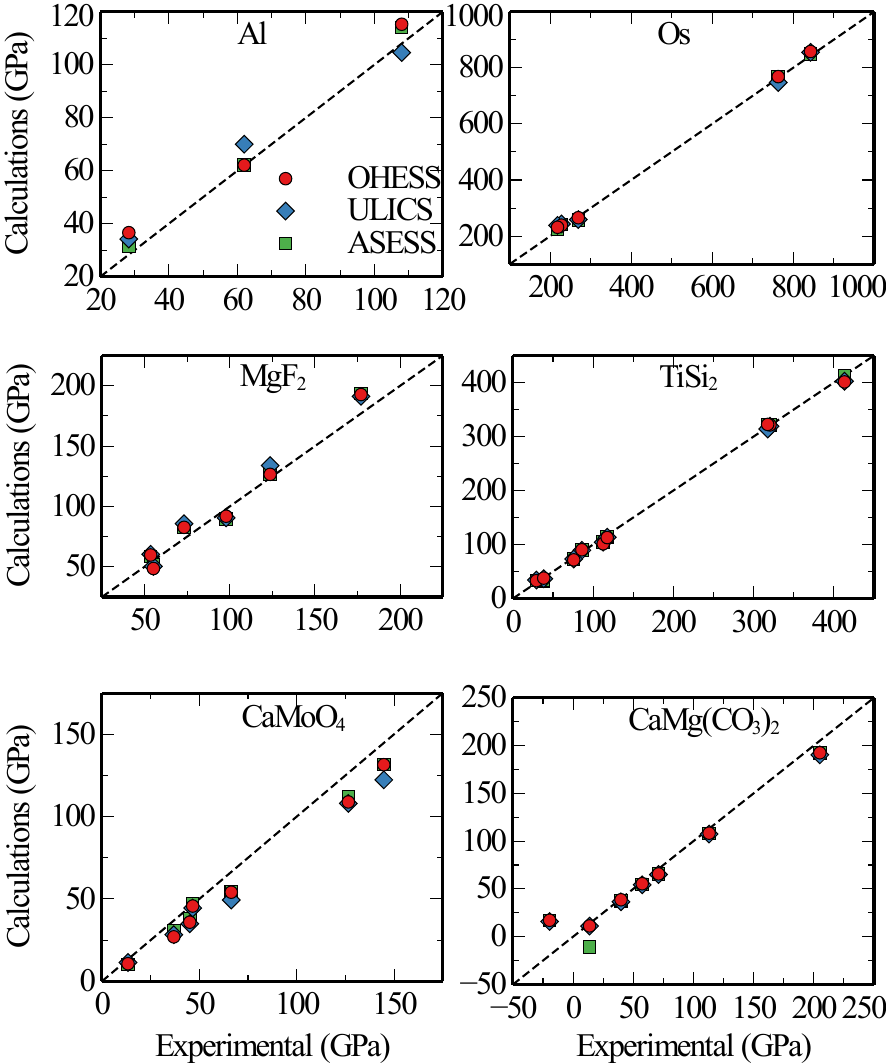}}
	\caption{Comparison of the elastic constants from OHESS, ASESS and ULICS
		with experiments\label{comparison}}
\end{figure}

Moving on to the hexagonal systems, we examined the precision of the OHESS, ASESS, and ULICS strain sets by calculating the elastic constants for materials such as hcp Os, Ti, and TiB$_2$. The derived values from our computations are provided in Table \ref{hcp}. For context, this table also includes pertinent data from the Materials Project ~\cite{Jong2015} and experiments~\cite{Pantea2009,Smithells1983,Spoor1997}. We observed a reasonable agreement between our computational findings and both the MP data and the referenced experimental values. For a more detailed insight, we offer a graphical comparison for Os in Fig.\ref{comparison}. These analyses collectively underscore the robustness and reliability of all three strain sets. Particularly for the hexagonal systems, the elastic constants computed using OHESS, ASESS, and ULICS are consistently in agreement with the experimentally observed values~\cite{Pantea2009,Smithells1983,Spoor1997}, underscoring their accuracy and applicability.

Our computations of the elastic constants for rhombohedral and tetragonal systems are detailed in Table \ref{tritet}. Within the realm of the trigonal system, we have materials such as Al$_2$O$_3$ and CaMg(CO$_3$)$_2$, which are classified under rhombohedral I and II crystal systems, respectively (Table \ref{num}). On the other hand, MgF$_2$ and CaMoO$_4$ are characterized within the tetragonal I and II crystal systems, respectively. When contrasting our findings obtained using the OHESS, ULICS, and ASESS strain sets with the corresponding data from the MP and select experimental studies \cite{Humbert1972,Alton1967,Smithells1983,Cutler1968}, we observe a consistent agreement. A point of note is the manifestation of negative values for certain elastic constants; these are attributable to the `-' Cartesian coordinate system as explained in Ref. \cite{Golesorkhtabar2013}. However, when focusing on the absolute magnitudes, these negative constants still show agreement with both MP and experimental observations.  For a more visual representation of our findings' accuracy, Fig. \ref{comparison} offers a comparative illustration of the elastic constants for materials like MgF$_2$, CaMoO$_4$, and CaMg(CO$_3$)$_2$. Overall, irrespective of the strain set employed—be it OHESS, ULICS, or ASES—our results exhibit close consistency and accuracy.

\begin{table*}[htp!]
	\caption{The calculated elastic constants of (Al$_2$O$_3$ and
	CaMg(CO$_3$)$_2$) and tetragonal (MgF$_2$ and CaMoO$_4$) systems compared
	with data from Materials Project and experiments.\label{tritet}}
\centering{
	\begin{tabular}{lllllllllll}
		\hline
		System & Method & $C_{11}$ & $C_{12}$ & $C_{13}$ & $C_{14}$ & $C_{15}$ &
		$C_{16}$ & $C_{33}$ & $C_{44}$ & $C_{66}$\\
		\hline
		CaMg(CO$_3$)$_2$ & OHESS & 192.4 & 65.7 & 55.4 & 17.2 & 11.3 &  & 108.3 &
		38.6 & \\
		& ULICS & 190.3 & 65.1 & 54.3 & 15.9 & 11.1 &  & 107.4 & 36.5 & \\
		& ASESS & 192.4 & 65.8 & 54.8 & 17.0 & -11.0 &  & 107.8 & 37.6 & \\
		& MP{\cite{Jong2015}} & 192 & 65 & 55 & 17 & 11 &  & 108 & 37 & \\
		& Exp.{\cite{Humbert1972}} & 205.0 & 71.0 & 57.4 & -19.5 & 13.7 &  &
		113.0 & 39.8 & \\
		CaMoO$_4$  & OHESS & 131.5 & 54.1 & 45.7 &  &  & 10.7 & 108.9 & 27.0 &
		35.9\\
		& ULICS & 122.3 & 49.4 & 44.5 &  &  & 11.4 & 108.0 & 28.3 & 34.9\\
		& ASESS & 131.5 & 54.1 & 46.9 &  &  & 10.1 & 112.2 & 30.5 & 37.9\\
		& MP{\cite{Jong2015}} & 133 & 56 & 47 &  &  & -12 & 113 & 29 & 37\\
		& Exp.{\cite{Alton1967}} & 144.7 & 66.4 & 46.6 &  &  & 13.4 & 126.5 &
		36.9 & 45.1\\
		Al$_2$O$_3$ & OHESS & 448.1 & 154.3 & 111.6 & 19.8 &  &  & 454.1 & 131.9 &
		\\
		& ULICS & 449.3 & 150.2 & 111.0 & 20.1 &  &  & 452.4 & 130.9 & \\
		& ASESS & 448.1 & 152.3 & 110.4 & 20.3 &  &  & 456.6 & 132.3 & \\
		& MP{\cite{Jong2015}} & 452 & 150 & 107 & 20 &  &  & 454 & 132 & \\
		& Exp.{\cite{Smithells1983}} & 497.4 & 164.0 & 112.2 & -23.6 &  &  &
		499.1 & 147.4 & \\
		MgF$_2$ & OHESS & 126.6 & 82.7 & 59.8 &  &  &  & 192.7 & 48.6 & 91.9\\
		& ULICS & 133.9 & 85.7 & 60.4 &  &  &  & 191.2 & 50.4 & 90.5\\
		& ASESS & 126.6 & 82.7 & 58.7 &  &  &  & 193.7 & 52.0 & 89.7\\
		& MP{\cite{Jong2015}} & 190 & 60 & 60 &  &  &  & 134 & 51 & 89\\
		& Exp.{\cite{Cutler1968}} & 123.7 & 73.2 & 53.6 &  &  &  & 177.0 & 55.2 &\\
		\hline
	\end{tabular}}

\end{table*}

For our examination of the orthorhombic and monoclinic crystal systems, we selected TiSi$_2$ and ZrO$_2$ as the representative materials, respectively. Our computed elastic constants for this class are detailed in Table~\ref{ortmon}. When juxtaposed with the MP data and experimental results, we obtained reasonable agreement for the orthorhombic TiSi$_2$, but our data deviates significantly for the monoclinic ZrO$_2$.  This discord in the results for ZrO$_2$ can possibly be traced back to inherent limitations in the VASP calculations. Specifically, when dealing with low symmetry intrinsic to the monoclinic structure, VASP might encounter challenges in delivering accurate stresses, thus leading to discrepancies in the resultant elastic constants.

\begin{table*}[htp!]
	\caption{The calculated elastic constants for the the orthorhombic TiSi$_2$
	and monoclinic ZrO$_2$ using OHESS, ASESS, and ULICS, in comparison with MP
	data and with experiments\label{ortmon}}

	\begin{tabular}{lllllllllllllll}
				\hline
		System & Method & $C_{11}$ & $C_{12}$ & $C_{13}$ & $C_{15}$ & $C_{22}$ &
		$C_{23}$ & $C_{25}$ & $C_{33}$ & $C_{35}$ & $C_{44}$ & $C_{46}$ & $C_{55}$
		& $C_{66}$\\
		\hline
		ZrO$_2$ & OHESS & 285.1 & 141.0 & 85.3 & 41.5 & 308.0 & 122.2 & -0.8 &
		225.6 & 2.9 & 65.2 & -6.2 & 79.7 & 112.2\\
		& ULICS & 328.5 & 140.3 & 111.0 & 36.0 & 313.0 & 125.7 & 13.5 & 270.0 &
		-5.1 & 55.0 & 0.43 & 72.1 & 104.1\\
		& ASESS & 285.1 & 141.0 & 86.3 & 42.1 & 307.6 & 123.5 & -0.5 & 231.6 &
		2.7 & 71.3 & -7.4 & 82.4 & 115.2\\
		& MP{\cite{Jong2015}} & 256 & 140 & 99 & -1 & 357 & 152 & 3 & 301 & -40 &
		114 & 8 & 80 & 70\\
		& Exp.{\cite{Chan1991}} & 361 & 142 & 55 & -21 & 408 & 196 & 31 & 258 &
		-18 & 100 & -23 & 81 & 126\\
		TiSi$_2$ & OHESS & 321.5 & 33.1 & 90.3 &  & 322.5 & 37.8 &  & 401.2 &  &
		71.6 &  & 101.4 & 112.9\\
		& ULICS & 319.1 & 34.2 & 89.4 &  & 313.9 & 36.6 &  & 402.5 &  & 73.1 &  &
		104.3 & 113.5\\
		& ASESS & 321.5 & 33.1 & 90.2 &  & 322.5 & 32.9 &  & 413.4 &  & 73.5 &  &
		104.8 & 114.4\\
		& MP{\cite{Jong2015}} & 310 & 31 & 90 &  & 389 & 25 &  & 307 &  & 72 &  &
		104 & 113\\
		& Exp.{\cite{Nakamura1993}} & 320.4 & 29.3 & 86.0 &  & 317.5 & 38.4 &  &
		413.2 &  & 75.8 &  & 112.5 & 117.5\\
				\hline
	\end{tabular}

\end{table*}

For our exploration into the triclinic crystal system—representing the lowest-symmetry 3D category—we selected ReS$_2$ as our prototype material to evaluate the computational accuracy of the three strain sets: OHESS, ULICS, and ASESS. The derived elastic constants of ReS$_2$ are presented in Table \ref{triclinic}. A notable observation here is the striking similarity between OHESS and ASESS within the context of the triclinic system. Indeed, they both employ an identical set of six strain matrices, as evidenced in Tables \ref{ohess} and \ref{asess}. Consequently, as clearly illustrated in Table \ref{triclinic}, the resultant elastic constants computed via OHESS and ASESS are identical. Delving deeper into a side-by-side analysis, we discern that the elastic constants derived from ULICS also agree well with their counterparts from both OHESS and ASESS.

\begin{table}[htp!]
	\caption{The calculated elastic constants for the triclinic ReS$_2$ using OHESS, ASESS, and ULICS methods.\label{triclinic}}

	\begin{tabular}{llccccccc}
		\hline
		 Method & $C_{11}$ & $C_{12}$ & $C_{13}$ & $C_{14}$ & $C_{15}$ &
		$C_{16}$ & $C_{22}$\\
		\hline
		 OHESS & 215.5 & 43.2 & 2.7 & -0.1 & 0.1 & 2.5 & 212.8\\
		 ULICS & 215.0 & 44.0 & 4.2 & -0.5 & -0.6 & 2.3 & 213.7\\
		 ASESS & 215.5 & 43.2 & 2.7 & -0.1 & 0.1 & 2.5 & 212.8\\
		  & $C_{23}$ & $C_{24}$ & $C_{25}$ & $C_{26}$ & $C_{33}$ & $C_{34}$ &
		$C_{35}$\\
		 OHESS & 2.5 & -0.1 & 0.0 & 2.3 & 6.9 & -0.1 & 0.0\\
		 ULICS & 4.3 & 0.4 & -0.3 & 1.9 & 8.0 & 0.2 & -0.3\\
		 ASESS & 2.5 & -0.1 & 0.0 & 2.3 & 6.9 & -0.1 & 0.0\\
		  & $C_{36}$ & $C_{44}$ & $C_{45}$ & $C_{46}$ & $C_{55}$ & $C_{56}$ &
		$C_{66}$\\
		 OHESS & 0.0 & 1.0 & -0.1 & 0.0 & 0.8 & 0.0 & 76.8\\
		 ULICS & -0.4 & 1.1 & -0.2 & -0.1 & 0.6 & 0.2 & 77.0\\
		 ASESS & 0.0 & 1.0 & -0.1 & 0.0 & 0.8 & 0.0 & 76.8\\
		\hline
	\end{tabular}

\end{table}

\subsubsection{2D materials}
Beyond 3D crystals, we extended our investigation to 2D crystal systems to assess the precision of elastic constant calculations employing the three strain sets: OHESS, ULICS, and ASESS. The outcomes of our computations are detailed in Table~\ref{2Dec}, juxtaposed against the theoretical values derived from other studies. Remarkably, our findings consistently agree with these previous results~\cite{Wang2015,Wang2021}. A point of emphasis is the consistent alignment observed between the results obtained using OHESS and those from both ULICS and ASESS. This further underscores the reliability and robustness of the OHESS methodology in diverse crystallographic systems.

\begin{table*}[htp!]
	\caption{The 2D in-plane elastic constants (N/m) of various 2D materials for OHESS, ULICS, and ASESS, respectively, compared with previous calculations.\label{2Dec}}
\centering{
		\begin{tabular}{lllllllll}
		\hline
		& $C_{11}$ &  & $C_{12}$ &  & $C_{22}$ &  & $C_{66}$ & \\
		Systems & Our work & Ref. & Our work & Ref. & Our work & Ref. & Our work &
		Ref.\\
		\hline
		Phosphorene & 103.4$^{[O]}$ & 105.2{\cite{Wang2015}} & 18.0$^{[O]}$ &
		18.4{\cite{Wang2015}} & 24.6$^{[O]}$ & 26.2{\cite{Wang2015}} &
		21.8$^{[O]}$ & 22.4{\cite{Wang2015}}\\
		& 110.6$^{[U]}$ & 104.4{\cite{Wang2021}} & 13.8$^{[U]}$ &
		21.6{\cite{Wang2021}} & 28.6$^{[U]}$ & 34.0{\cite{Wang2021}} &
		24.3$^{[U]}$ & 27.4{\cite{Wang2021}}\\
		& 104.1$^{[A]}$ &  & 17.4$^{[A]}$ &  & 24.6$^{[A]}$ &  & 22.8$^{[A]}$ &
		\\
		AuSe & 34.1$^{[O]}$ &  & 2.7$^{[O]}$ &  & 9.3$^{[O]}$ &  & 3.3$^{[O]}$ &
		\\
		& 34.9$^{[U]}$ &  & 2.5$^{[U]}$ &  & 9.4$^{[U]}$ &  & 3.3$^{[U]}$ & \\
		& 34.2$^{[A]}$ &  & 2.7$^{[A]}$ &  & 9.3$^{[A]}$ &  & 3.3$^{[A]}$ & \\
		Graphene & 353.2$^{[O]}$ & 358.1{\cite{Wang2015}} & 63.7$^{[O]}$ &
		60.4{\cite{Wang2015}} &  &  &  & \\
		& 353.2$^{[U]}$ & 349.1{\cite{Wang2021}} & 64.2$^{[U]}$ &
		60.3{\cite{Wang2021}} &  &  &  & \\
		& 353.2$^{[A]}$ &  & 63.9$^{[A]}$ &  &  &  &  & \\
		MoS$_2$ & 136.9$^{[O]}$ & 131.4{\cite{Wang2015}} & 33.1$^{[O]}$ &
		32.6$^{}${\cite{Wang2015}} &  &  &  & \\
		& 137.1$^{[U]}$ & 128.9{\cite{Wang2021}} & 33.3$^{[U]}$ &
		32.6{\cite{Wang2021}} &  &  &  & \\
		& 136.9$^{[A]}$ &  & 33.7$^{[A]}$ &  &  &  &  & \\
		FeSe & 58.2$^{[O]}$ &  & 22.7$^{[O]}$ &  & 38.1$^{[O]}$ &  &  & \\
		& 57.6$^{[U]}$ &  & 22.5$^{[U]}$ &  & 38.3$^{[U]}$ &  &  & \\
		& 58.4$^{[A]}$ &  & 22.3$^{[A]}$ &  & 38.2$^{[A]}$ &  &  & \\
		\hline
	\end{tabular}}

\end{table*}

\begin{figure}[htp!]
	\resizebox{8.5cm}{!}{\includegraphics{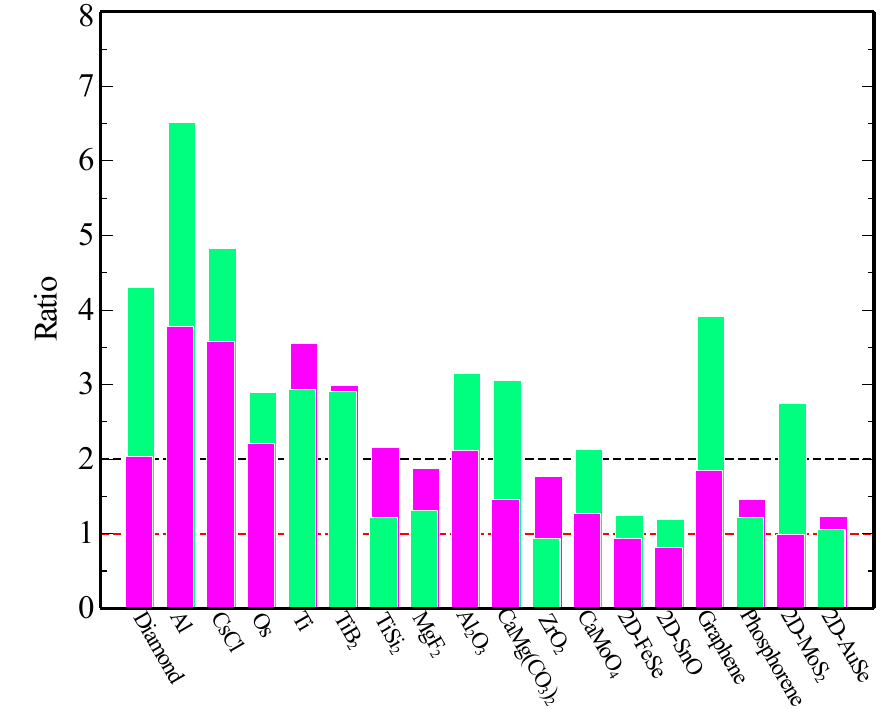}}
	\caption{The ratio of the total computational time used by ULICS (pink) and ASESS (green) to that used by OHESS for calculating elastic constants of different crystals in 3D and 2D systems. \label{time-fig}}
\end{figure}

\subsection{Computation efficiency of different strain sets}\label{IIIB}
While the OHESS, ULICS, and ASESS methodologies exhibit comparable accuracy, it is pertinent to assess the computational efficiency of each, especially when calculating elastic constants. Specifically, the duration each method requires to complete these calculations can serve as a metric of efficiency. To provide context, all the calculations for elastic constants were executed on a specific hardware platform. This platform was equipped with two Xeon E5-2683 CPUs, offering a combined processing power of 2.0 GHz distributed across 28 cores. We recorded the time each methodology used for every calculation, both for 3D and 2D elastic constants. These timings are presented in Table \ref{time}. An examination of the data in Table \ref{time} reveals pronounced differences in the time each methodology requires. It's noteworthy that, despite all three methods — OHESS, ULICS, and ASESS — achieving comparable accuracies, the time they take to compute the elastic constants is markedly different. This suggests that while accuracy is a shared strength among the three methods, computational efficiency varies, and it's an aspect that should be considered when selecting a method for specific applications.

\begin{table*}[htp!]
	\caption{The computation time (in seconds) used by the three types of strain sets, OHESS, ULICS and ASESS.\label{time}}
\centering{
	\begin{tabular}{cccccccc}
		\hline
		Dimensional & Strain sets & Diamond & Al & CsCl & Os & Ti & TiSi$_2$\\
		\hline
		3D & OHESS & 222 & 392 & 196 & 555 & 218 & 2317\\
		& ULICS & 452 & 1483 & 701 & 1225 & 773 & 4993\\
		& ASESS & 956 & 2556 & 946 & 1605 & 640 & 2811\\
		&  & MgF$_2$ & Al$_2$O$_3$ & CaMg(CO$_3$)$_2$ & ZrO$_2$ & TiB$_2$ &
		CaMoO$_4$\\

		& OHESS & 3580 & 7074 & 22200 & 27474 & 401 & 17207\\
		& ULICS & 6720 & 14958 & 32385 & 48593 & 1196 & 21955\\
		& ASESS & 4687 & 22260 & 67936 & 25703 & 1165 & 36681\\
		2D &  & ReS$_2$ & 2D-FeSe & Graphene & Phosphorene & 2D-MoS$_2$ &
		2D-AuSe\\

		& OHESS & 59701 & 3898 & 404 & 5396 & 2298 & 17860\\
		& ULICS & 80100 & 3662 & 749 & 7858 & 2278 & 21916\\
		& ASESS & 59701 & 4847 & 1580 & 6545 & 6311 & 18974\\
		\hline
	\end{tabular}}

\end{table*}

To offer a comprehensive assessment of the computational efficiency among the three strain sets – ULICS, ASESS, and OHESS – we graphically represented the time ratios. Specifically, we juxtaposed the times taken by ULICS and ASESS against that of OHESS, as presented in Figure\ref{time-fig}. This visual representation underscored a noteworthy observation: OHESS consistently emerged as the most computationally efficient strain set. Digging deeper into the data for 3D crystal systems, OHESS exhibited more than double the efficiency compared to both ULICS and ASESS in the majority of cases. However, there were exceptions. In instances involving graphene and 2D-MoS$_2$, OHESS's efficiency was on par with that of ULICS and ASESS. A closer look at the inherent characteristics of each strain set offers some clarity. For instance, ASESS always incorporates six strain sets for 3D systems and three for 2D, as detailed in Table~\ref{asess}. This intrinsic attribute could contribute to its comparatively reduced efficiency. In contrast, both OHESS and ULICS, for the most part, employ a similar number of strain sets. An exception is observed in the monoclinic crystal system, where their numbers diverge. Furthermore, their strain set counts are identical for every crystal system except the triclinic. The reduced efficiency of ULICS can be attributed to its approach that diminishes the crystal symmetry through the utilization of coupled strain sets. This approach subsequently increases the calculation time required for the stress calculations.

\subsection{Discussions}
In our extensive assessment, we rigorously compared the three distinct strain matrix sets - OHESS, ULICS, and ASESS - in terms of their accuracy and computational efficiency for calculating elastic constants across both 3D and 2D systems. Our findings indicate that OHESS considerably speeds up the calculation process for second-order elastic constants. For 3D systems, our evaluation encompassed a broad spectrum of crystal structures, spanning from cubic to triclinic types. Meanwhile, our 2D tests were focused on square, hexagonal, and rectangular lattices. Despite all three strain matrix sets demonstrating comparable accuracy in deriving elastic constants, OHESS stands out with superior computational efficiency. This distinction becomes particularly evident in crystal structures characterized by high symmetry, such as the cubic and hexagonal systems (as illustrated by examples like Diamond, Al, CsCl, Os, Ti, and TiB$_2$, all highlighted in Figure~\ref{time-fig}). However, the efficiency of OHESS tends to diminish in comparison to ULICS and ASESS as the symmetry of the crystal system decreases. Two primary factors contribute to this observation. Firstly, the test cases inherently possess lower symmetries before the application of either the OHESS or ULICS strain sets. Consequently, the differences post-application are not as pronounced as they are with higher symmetry crystal structures. Secondly, the number of strain sets employed in the OHESS is similar to that in the ASESS approach, causing the efficiency gap between OHESS and the other two to reduce marginally. Shifting our focus to 2D systems, OHESS's computational efficiency advantage over ULICS is less pronounced. The primary reason for this is the inherently lower symmetry of the original 2D lattice. Post-strain application, OHESS maintains a symmetry closely aligned with ULICS. Nonetheless, OHESS outperforms ASESS in certain scenarios, such as with graphene and 2D MoS$_2$. This superior efficiency stems from the fewer strain sets incorporated in the OHESS method.

\section{Conclusions}\label{conclusion}
In our study, we introduced the optimized high-efficient strain-matrix set (OHESS) as an innovative approach to expedite the calculation of elastic constants of materials based on the foundational stress-strain relationship delineated in Hooke's law. Following an exhaustive evaluation, comparing OHESS with other strain-matrix sets like ULICS and ASESS in terms of computational accuracy and efficiency, we determined that OHESS stands out as the most efficient method for calculating elastic constants across both 3D and 2D scenarios without compromising on accuracy. This breakthrough implies that OHESS has the potential to significantly enhance the speed of elastic constant calculations using stress-strain relations moving forward. Such advancements are invaluable for swiftly assessing the stability of novel crystal structures in contemporary material design, especially when it comes to populating databases dedicated to elastic constants, such as the Materials Project elastic constant database.

\section{Acknowledgments}

We acknowledge the support from the National Natural Science Foundation of China (41574076, 11974091), the Key Research Scheme of Henan Universities (18A140024), and the Research Scheme of LYNU Innovative Team under Grant No. B20141679.

\section{References}
\bibliographystyle{elsarticle-num}

\end{document}